\documentclass{llncs}

\usepackage{time}
\usepackage{color}
\usepackage{pslatex}
\usepackage{pifont}
\usepackage{epsfig}
\usepackage{psfig}
\usepackage{graphics}
\usepackage{float}
\usepackage{algorithm}
\usepackage{algorithmicx}     
\usepackage{algcompatible}    
\usepackage{alltt}            
\usepackage{latexsym}         
\newcommand{\algorithmicinput}{\textbf{Input:~}}
\newcommand{\algorithmicoutput}{\textbf{Output:~}}

\newcommand{\vars}{\cal{V}}
\newcommand{\functors}{\cal{F}}
\newcommand{\neck}{\texttt{:-}}
\newcommand{\entails}{\vdash}
\newcommand{\myapprox}{\leadsto}
\newcommand{\mycup}{\hbox{
\begin{picture}(1,1)(1,1)
\put(2,1){$\cdot$}
\put(0,0){$\cup$}
\end{picture}
}\ }
\newcommand{\mybigcup}{\hbox{
\begin{picture}(1,1)(1,1)
\put(2,1){$\cdot$}
\put(0,0){$\bigcup$}
\end{picture}
}\ }

\newcommand{\comment}[1]{}

\newcounter{mnotei}
\newcommand{\ciao}{\texttt{Ciao}}

\begin{document}
\title{  Towards Parameterized Regular Type Inference\\
         Using Set Constraints }

 \author{\sc
        F.Bueno,\inst{1} 
        J.Navas,\inst{2} and
        M.Hermenegildo\inst{1,3}}

\institute{Technical U. of Madrid (Spain)
           \and National University of Singapore (Republic of Singapore)
           \and IMDEA Software (Spain)}

\maketitle

\begin{abstract}
  We propose a method for inferring \emph{parameterized regular types}
  for logic programs as solutions for systems of constraints over sets
  of finite ground Herbrand terms (set constraint systems).  Such
  parameterized regular types generalize \emph{parametric} regular
    types by extending the scope of the parameters in the type
  definitions so that such parameters can relate the types of
  different predicates. We propose a number of enhancements to the
  procedure for solving the constraint systems  
  that improve the precision of the type
  descriptions inferred. The resulting algorithm, together with
  a procedure to establish a set constraint system from a logic
  program, yields a program analysis that infers tighter safe
  approximations of the success types of the program than previous
  comparable work, offering a new and useful efficiency vs.\ precision 
  trade-off.  This is supported by experimental results, 
  which show the feasibility of our analysis.

\end{abstract}

\section{Introduction}
\label{sec:intro}

Type inference of logic programs is the problem of computing, at
compile time, a representation of the terms that the predicate
arguments will be bound to during execution of the program. This kind
of type inference involves not only assigning types to procedure
arguments out of a predefined set of type definitions, as in
traditional type inference, but also the more complex problem of
inferring the type definitions themselves, similarly to what is done
in \emph{shape analysis}.
Although most logic programming languages are either untyped or allow
mixing typed and untyped code, inferring type information for the
entire program is important since it allows the compiler to generate
more efficient code and it has well-known advantages in detecting
programming errors early.
For instance, simple uses of such information include better indexing,
specialized unification, and more efficient code generation.  More
advanced uses include compile-time garbage collection and
non-termination detection. 
There are also other areas in which type information can be
useful. For example, during verification and debugging it can provide
information to the programmer that is not straightforward to obtain by
manual inspection of the program.

In this paper we use the \emph{set constraint-based}
approach~\cite{Heintze92,Heintze92PhD}. We propose 
an algorithm for solving a set constraint system that relates the set
of possible values of the program variables, by transforming it into a
system whose solutions provide the type definitions 
for the variables involved in \emph{parameterized regular form}.
We focus on types which are conservative approximations of the
meaning of predicates, and hence, over-approximations
of the \emph{success set} (in contrast to the approach of 
inferring well-typings, as in, e.g.,~\cite{BGV05,SBG08}, 
which may differ from the
actual success set of the program). 
Type inference via set constraint solving was already proposed
in~\cite{HeintzeJaffar90,Heintze92PhD}. However, most existing
algorithms~\cite{HeintzeJ92b,Heintze92,set-based-absint-padl-short}
are either too complicated or lacking in precision for some classes of
programs. We try to alleviate these problems by, first, generating
simple equations; second, using a comparatively straightforward
procedure for solving them; and, third, using a non-standard operation
during solving that improves precision by ``guessing'' values. At the
same time, we attack a more ambitious objective since our resulting
types, that we call \emph{parameterized}, are more expressive than in
previous proposals.

{\small
\begin{figure}[t]
\begin{center}
\begin{tabular}{|l|}
\hline
   \begin{tabular}{|l|l|}
   \begin{minipage}[t]{0.4\textwidth}
   \vspace{-4\baselineskip}
   \textsf{(a) Program}:
   \vspace{-0.8\baselineskip}
\begin{verbatim}

:- typing append(A1,A2,A3).

append([],L,L).
append([X|Xs],Ys,[X|Zs]):- 
       append(Xs,Ys,Zs).
\end{verbatim}
   \end{minipage} 
   &
   \begin{minipage}[t]{0.5\textwidth}
      \begin{tabular}{l}
      \begin{minipage}[t]{1.0\textwidth}
      \textsf{(b) Parametric regular types}:
      \vspace{-0.8\baselineskip}
\begin{verbatim}
:- type A1(X)   -> [] | [X|A1(X)].
:- type A2(W)   -> W.
:- type A3(Y,Z) ->  Z | [Y|A3(Y,Z)].
\end{verbatim}
      \end{minipage} 
      \\ 
      \hline
      \begin{minipage}[t]{1.0\textwidth}
      \textsf{(c) Parameterized regular types}:
      \vspace{-0.8\baselineskip}
\begin{verbatim}
:- type A1 -> [] | [X|A1].
:- type A2 ->  .
:- type A3 -> A2 | [X|A3].
\end{verbatim}
      \end{minipage}
      \end{tabular}
   \end{minipage}
   \end{tabular}
\\
\hline
   \begin{tabular}{|l|l|}
   \begin{minipage}[t]{0.55\textwidth}
   \textsf{(d) Local parameters}:
   \vspace{-0.8\baselineskip}
\begin{verbatim}
:- typing append(A1(E),A2(T),A3(E,T)).
:- typing nrev(N1(E),N2(E)),
          append(A1(E),A2([]),A3(E,[])).
\end{verbatim}
   \end{minipage} 
   &
   \begin{minipage}[t]{0.35\textwidth}
   \textsf{(e) Global parameters}:
   \vspace{-0.8\baselineskip}
\begin{verbatim}
:- typing nrev(R1,R2).
:- type R1 -> [] | [X|R1].
:- type R2 -> [] | [X|R2].
\end{verbatim}
   \end{minipage}
   \end{tabular}
\\
\hline
\end{tabular}
\end{center}
\vspace{-1.5\baselineskip}
\caption{Parametric vs parameterized regular types}
\label{fig:motivation}
\vspace{-\baselineskip}
\end{figure}
}

Consider for instance the \texttt{append/3} program in
Fig.~\ref{fig:motivation}.a, and assume the type descriptors
\texttt{A1}, \texttt{A2}, and \texttt{A3} for each predicate
argument. Most state-of-the-art analyses will simply infer that the
first argument (type \texttt{A1}) is a list, leaving open the types of
the other two arguments.  In Fig.~\ref{fig:motivation}.b we show
classical parametric types for \texttt{append/3}. We are unaware of an
existing proposal that is able to infer them as an approximation to
the success set of the predicate
(see Sec.~\ref{sec:related}). Even if we had an analysis that inferred
them, they are still less expressive (or need more elaboration for the
same expressiveness) than our proposal, as we will discuss.

The parametric types in Fig.~\ref{fig:motivation}.b denote the expected list
type for \texttt{A1} which is parametric on some \texttt{X}. Note that
\texttt{A2} is unbound since it may be instantiated to any term. On
the other hand, \texttt{A3} is an open-ended list of elements of some type
\texttt{Y} whose tail is of some type \texttt{Z}.
A key observation is that while there is a clear relation between the type
of the elements of \texttt{A1} and \texttt{A3}, and between the type of the
tails in \texttt{A2} and \texttt{A3}, these relations are not captured. 
In Fig.~\ref{fig:motivation}.c, we show a
desirable, more accurate type for \texttt{append/3}. It denotes that
the type of \texttt{A3} is that of open-ended lists of elements of the
same type as \texttt{A1} with tail of the same type as \texttt{A2}. 

The relations between the types of arguments can be captured with
parametric types only if type parameters are instantiated. For
example, the first typing in Fig.~\ref{fig:motivation}.d captures the
desired relations for \texttt{append/3}.  Instead, by using global
type parameters, as we propose, the parametric type definitions may
exhibit the required exact relations right from the inference of the
type definitions alone.  Although the absence of typings to express
the same relations is a small advantage, a bigger one might be
expected with regard to the analysis.  By using parametric types, an
analysis should not only infer type definitions but also typings
showing the exact values for the type parameters. It is not clear at
all how this might be done (note that typings are not the types of
calls). Our proposal infers type definitions alone, as usual, and yet
are at least as expressive as parametric types \emph{with} typings.

Consider a program construct where two different predicates
share a variable, so that the corresponding arguments have the same type.
This happens, for example, with the arguments of \texttt{nrev/2}
and \texttt{append/3} in the classical naive reverse program (see
Ex.~\ref{ex:initial}). This property cannot be captured with
(standard) typings for the predicates if parametric type definitions
are used. One would need something like the second typing in
Fig.~\ref{fig:motivation}.d. However, this is not usually a valid typing
(it types several predicate atoms at the same time), and it is also 
not intuitive how it could be inferred.
In contrast, the parameterized type definition of 
Fig.~\ref{fig:motivation}.e, together with those of (c),
easily capture the property, by sharing the type variable global
parameter \texttt{X}.
Therefore, more precise types (called \emph{parameterized
  regular types}~\cite{Mishra85}) can be produced if the scope of each
type variable in a type definition is broader than the definition
itself, so that the types of different arguments can be related. 

As discussed in detail in Sec.~\ref{sec:related}, we believe that no
previous proposal exists for inferring types for logic programs with
the expressive power of parameterized regular types. In addition, our
proposal fits naturally in the set constraint-based approach. In this
context we also define a number of enhancements to the solving
procedure.  The result is a simple but powerful type inference
analysis. Our preliminary experiments also show that our analysis runs
in a reasonable amount of time.

\section{Preliminaries}
\label{sec:preli}

Let $V$ be a non-terminal which ranges over (set) variables $\vars$, and
let $f$ range over functors (constructors) $\functors$ of given arity $n\geq 0$.
\emph{Set expressions}, set expressions in \emph{regular form}, and in 
\emph{parameterized form} are given by $E$ in the following grammars:
\[\begin{array}{llcclccl}
  \mbox{Expressions: } & 
     \multicolumn{7}{l}{
     E ::= \emptyset ~|~ V ~|~ f(E_1,\ldots,E_n) ~|~ 
           E_1\cup E_2 ~|~ E_1\cap E_2 } \\
  \mbox{Regular form: } & E ::= \emptyset ~|~ N &&&
     \multicolumn{4}{l}{
     N ::= V ~|~ f(V_1,\ldots,V_n) ~|~ N_1\cup N_2 } \\
  \mbox{Parameterized form: } & E ::= \emptyset ~|~ N 
   &&& N ::= N_1\cup N_2 ~|~ R 
   &&& R ::= V ~|~ f(V_1,\ldots,V_n) ~|~ V\cap R
\end{array}
\]

The meaning of a set expression is a set of (ground, finite) terms, and
is given by the following semantic function $\mu$ under an assignment
$\sigma$ from variables to sets of terms:
\[\begin{array}{rclcccrcl}
  \mu(\emptyset,\sigma) & = & \emptyset &&&&
  \mu(V,\sigma) & = & \sigma(V) \\
  \mu(E_1\cup E_2,\sigma) & = & \mu(E_1,\sigma)\cup\mu(E_2,\sigma) &&&&
  \mu(E_1\cap E_2,\sigma) & = & \mu(E_1,\sigma)\cap\mu(E_2,\sigma) \\
  \multicolumn{4}{r}{
  \mu(f(E_1,\ldots,E_n),\sigma) } & = &
  \multicolumn{4}{l}{
       \{ f(t_1,\ldots,t_n) ~|~ t_i\in\mu(E_i,\sigma) \} }
\end{array}
\]

Let $E_{1}$ and $E_{2}$ be two set expressions, then a \emph{set
  equation} (or equation, for short) is of the form $E_1 = E_2$. A
\emph{set equation system} is a set of set equations.  A solution
$\sigma$ of a system of equations $S$ is an assignment that maps
variables to sets of terms which satisfies:
$\mu(e_1,\sigma) = \mu(e_2,\sigma) \mbox{ for all } (e_1=e_2)\in S$.
We will write $S\entails e_1=e_2$ iff every solution of $S$ is a solution
of $e_1=e_2$.
A set equation system is in \emph{top-level form} if in all
expressions of the form $f(x_1,\ldots,x_n)$, all the $x_i$ are
variables.
The \emph{top-level} variables of a set expression are the variables
which occur outside the scope of a constructor.

A \emph{standard set equation system} is one in which all
equations are of the form $V=E$ (i.e., lhs are variables and rhs set
expressions) and there are no two equations with the same
lhs. Equations where the rhs is also a variable will be called {\em
  aliases}.
In a standard set equation system variables which are not in the
lhs of any equation are called \emph{free variables} (since they are
not constrained to any particular value). Variables which do appear in
lhs are called non-free.

A \emph{regular set equation system} is one which is standard, all rhs
are in regular form, has no top-level variables except for aliases, and 
also no free variables.
A regular set equation system is in direct syntactic correspondence
with a set of regular type definitions. \emph{Regular types} are
equivalent to regular term grammars where the type definitions are the
grammar rules. In a regular set equation system the set
variables act as the type symbols, and each equation of the form
$x=e_1\cup\ldots\cup e_n$ acts as $n$ grammar rules of the form
$x::=e_j$, $1\leq j\leq n$.
By generalizing regular set equation systems to allow free variables
what we obtain is the possibility of having parameters within the
regular type definitions. However, when free variables are allowed
intersection has to be allowed too: given that free variables are not
constrained to any particular value, intersections cannot be
``computed out.'' 

A \emph{leaf-linear set equation system}~\cite{Mishra85} is one which
is standard, all rhs are in parameterized form, and
all top-level variables are free.
Note that leaf-linear set equation systems are the minimal extension
of regular equation systems in the above mentioned direction, in the sense
that intersections are reduced to a minimal expression: several free
variables and only one (if any) constructor expression.
More importantly, a leaf-linear set equation system is more expressive
than parametric regular types, since parameters have the scope of the
whole system, instead of the particular type definition in which they
occur, as in parametric type definitions.

\section{Type inference}
\label{sec:solve}

In this section, we present the different components of our analysis
method for inferring parameterized regular types. 
First, a set equation system is derived from the syntax of the
program, then, the system is solved, and finally it is projected onto
the program variables 
providing the type definitions for such variables.
The resulting equation system is in \emph{solved form}, i.e., it is
leaf-linear. Such a system will be considered a fair representation of the
solution to the original set equation system, since it denotes a set of
parameterized regular types.
When the system is reduced to solved form, the parameters of such a solution
are the free variables.

\paragraph{\textbf{Generating a set equation system for a program}.}
Let $P$ be a program, $\Pi_P$ the set of predicate symbols in
$P$, ranked by their arity, and $P|_p$ 
the set of rules defining predicate $p$ in program $P$.
Our analysis assumes that
all rules in $P$ have been \emph{renamed} apart so that they do
  not have variables in common.
For each $p \in \Pi_{P}$ of arity $n$ we associate a
\emph{signature} of $p$ defined as $\Sigma(p) = p(x_1,\ldots,x_n)$
where $\{x_1,\ldots,x_n\}$ is an ordered set of $n$ new variables,
one for each argument of $p$.
For atom $A$, let $[A]_j$ denote its $j$-th argument. For a predicate 
$p \in \Pi_P$ with arity $n$ we define $C_p$ and the
initial set equation system, $E$, for $P$ as follows.
In order to avoid overloading symbol $\cup$, to clarify the presentation,
we will use $\mycup$ for the usual set union, while $\cup$ will stand
for the symbol occurring in set expressions.

\begin{equation}
\label{eq:gen}
\begin{array}{lccccl}
E = \mybigcup\{ C_p ~|~ p\in \Pi_P \} &&&&
C_{p} = C_{Head} ~\mycup~ C_{Body}
\end{array}
\end{equation}
\[\begin{array}{lclcl}
 C_{Head} & = & \{\ x_{j} ~ = ~ \bigcup \{ [H]_j ~|~ (H \neck B) \in P|_p \} 
              \ ~|~ x_j \in vars(\Sigma(p)) \ \} \\
 C_{Body} & = & \{\ y ~=~ \bigcap\{ [\Sigma(A)]_{i} ~|~ [A]_{i} = y,~ A \in B \}
              \ ~|~ (H \neck B) \in P|_p, \  y \in vars(B) \ \} \\
          & \mycup & \{\ y ~=~ [\Sigma(A)]_{i}\cap t ~|~ [A]_{i} = t,~ A \in B,~
            (H \neck B) \in P|_p,\ t \not\in vars(B),\ y \mbox{ fresh var} \ \}
\end{array}
\]

\begin{example}\label{ex:initial}
  Take signatures \texttt{append(A1,A2,A3)} and \texttt{nrev(N1,N2)} in the 
  following program for naive reverse.
  The equation system for \texttt{nrev/2} is 
  $C_{nrev/2} ~~=~~ C_{H} \mycup\ C_{B}$.
\begin{verbatim}
nrev([],[]).               
nrev([X|Xs],Ys):- nrev(Xs,Zs), append(Zs,[X],Ys).
\end{verbatim}
\vspace{-0.5\baselineskip}
  \[
  \begin{array}{lclcl}
  C_{H} & = & \{~ \texttt{N1 = []} \cup \texttt{[X|Xs]},~ 
                 \texttt{N2 = []} \cup \texttt{Ys} ~\} \\
  C_{B} & = & \{~ \texttt{Xs = N1}, ~\texttt{Ys = A3},~ 
                 \texttt{Zs = N2} \cap \texttt{A1},~
                 \texttt{W = [X]} \cap \texttt{A2} ~\}
  \end{array}
  \]
\end{example}

Note that a system $E$ which results from Eq.~\ref{eq:gen}
is in standard form. Moreover, to put it also in top-level form
we only need to repeatedly rewrite every subexpression of every equation
of $E$ of the form $f(e_1,\ldots,e_j,\ldots,e_n)$ into 
$f(e_1,\ldots,y_j,\ldots,e_n)$, whenever $e_j$ is not a variable, 
adding to $E$ equation $y_j=e_j$, with $y_j$ a new fresh variable, 
until no further rewriting is possible. The new equations added to 
$E$ are in turn also rewritten in the same process.
We call the resulting system $Eq(P)$. Obviously, $Eq(P)$ is equivalent
to $E$ in Eq.~\ref{eq:gen}.
%

\paragraph{\textbf{Analysis of the program}.}
In order to analyze a program and infer its types, we follow the
call graph of the program bottom-up, as explained in the following.
First, the call graph of the program is built and its strongly connected 
components analyzed. Nodes in the same component are replaced by a single
node, which corresponds to the set of predicates in the original nodes.
The (incoming or outgoing) edges in the original nodes are now edges of the 
new node.
The new graph is partitioned into levels. The first level consists of
the nodes which do not have outgoing edges. Each successive level
consists of nodes which have outgoing edges only to nodes of lower
levels.
The analysis procedure processes each level in turn, starting from the
first level. Predicates in the level being processed can be analyzed
one at a time or all at once.

Each graph level is a subprogram of the original program. To analyze
a level, equations are set up for its predicates as by Eq.~\ref{eq:gen},
and copies of the solutions already obtained for the predicates in lower
levels are added. To do this, the signatures and solutions of the 
predicates in lower levels are renamed apart. The new signatures replace 
the old ones when used in building the equations for the subprogram. 
The new copies of the solutions are added to the set equation system.
For a given predicate, there is a different copy for each atom of that
predicate which occurs in the subprogram being analyzed.

\begin{example}\label{ex:initial-sccs}
  Consider the following (over-complicated) contrived predicate to
  declare two lists identical, which   calls predicate \texttt{nrev/2}
  of Ex.~\ref{ex:initial} twice.
\begin{verbatim}
same(L1,L2):- nrev(L1,L), nrev(L,L2).
\end{verbatim}
  Predicate \texttt{nrev/2}, which is in a lower scc, would have been analyzed
  first. Taking   signatures \texttt{same(S1,S2)} and \texttt{nrev(N1,N2)}, the
  solution for \texttt{nrev/2} would be:
  \[ \{\ \texttt{N1} = \texttt{[]} \cup \texttt{[X|N1]},\
         \texttt{N2} = \texttt{[]} \cup \texttt{[X|N2]}\ \}
  \]
  Since there are two calls to this predicate in the above definition
  for \texttt{same/2}, two copies of the above equations (renaming the
  above equations for $\langle \texttt{N1},\texttt{N2} \rangle$ into
  $\langle \texttt{N11},\texttt{N12} \rangle$ and $\langle
  \texttt{N21},\texttt{N22} \rangle$) would be added.  The initial
  equation system for \texttt{same/2} is thus:
\vspace{-0.5\baselineskip}
  \[
  \begin{array}{ccc}
  \{~ & \texttt{S1 = L1},~ \texttt{S2 = L2},~
        \texttt{L1 = N11},~ \texttt{L2 = N22},~
        \texttt{L = N12} \cap \texttt{N21},~ \\
      & \texttt{N11} = \texttt{[]} \cup \texttt{[X1|N11]},\
        \texttt{N21} = \texttt{[]} \cup \texttt{[X2|N21]},\ \\
      & \texttt{N12} = \texttt{[]} \cup \texttt{[X1|N12]},\
        \texttt{N22} = \texttt{[]} \cup \texttt{[X2|N22]}\ 
      & \}
  \end{array}
  \]
  Note that the two arguments of \texttt{same/2} have, in principle,
  list elements of different type (\texttt{X1} and \texttt{X2}). That
  they are in fact of the same type will be recovered during the solving
  of the equations, in particular, the equation for \texttt{L}, the rhs
  of which is an intersection (unification). This will be done by an
  operation we propose (BIND) which will make an appropriate ``guess'',
  as explained later.
\end{example}

Note that the copies are required because the type variables involved
might get new constraints during analysis of the subprogram. Such
constraints are valid only for the particular atom related to the
given copy.
Also, as the example above shows, using a single copy would
impose constraints, because of the sharing of the same parameter between
equations referring to the same and single copy (as it would have been
the case with \texttt{X} in the example, had we not copied the solution
for \texttt{nrev/2}). Such constraints might be, in principle, not true
(though in the example it is finally true that both lists have elements
of the same type).
Thus, copies represent the types of the program atoms occurring at the
different program points (i.e., different call patterns).

Note also that equations with intersection in the expression in the rhs
will be used to capture unification, as it is the case also in the
example. This occurs sometimes by simplification of intersection
(procedure SIMP below), but most of the times from operation BIND,
that mimics unification. Such equations are particularly useful. This
is also true even if the equation is such that its lhs is a variable
which does not occur anywhere else in the program (as it is the case
of \texttt{W} in Ex.~\ref{ex:initial}).
Additionally, in these cases, if the rhs finally becomes the empty set,
denoting a failure, this needs to be propagated explicitly, since the
lhs is a variable not related to the rest of the equations.

The details of the solving procedure are explained below. The procedure
is based on the usual method of treating one equation at a time,
the lhs of which is a variable, and replacing every (top-level) occurrence
of the variable by the expression in the rhs everywhere else.
Once the system is set up as explained, it is first normalized and then
solved as described in the following.

\paragraph{\textbf{Normal form}.}
The normal form used is \emph{Disjunctive Normal Form (DNF)}, 
plus some simplifications based on equality
axioms for $\cap$, $\cup$, and $\emptyset$, which transform expressions
into parameterized form. Once the system is in top-level form, two
auxiliary algorithms are used to
rewrite equations to achieve normal form. These algorithms are
based on semantic equivalences, and therefore preserve solutions.

The first algorithm, DNF, puts set expressions in a set equation in
disjunctive normal form. Note that if all expressions of an equation
$q$ are in top-level form then those of equation $DNF(q)$ are in
parameterized form, except for nested occurrences of $\emptyset$ and
possibly several occurrences of constructor expressions in
conjuncts. This is taken care of in the second algorithm, SIMP.

SIMP simplifies set expressions in an equation system $E$ by
repeatedly rewriting every subexpression based on several equivalences
until no further rewriting is possible, as follows:
\begin{enumerate}
\item $e \cap \emptyset$ \label{eq:emptycap}
      $\myapprox$ $\emptyset$ \hfill {\sc inter.~simplification}
\item $e\cap e$ \label{eq:absorbcap} 
      $\myapprox$ $e$ \hfill {\sc inter.~absorption}
\item $e \cup\emptyset$ \label{eq:emptycup}
      $\myapprox$ $e$ \hfill {\sc union simplification}
\item $e \cup e$ \label{eq:absorbcup}
      $\myapprox$ $e$ \hfill {\sc union absorption}
\item $e_1 \cup (e_1 \cap e_2)$ \label{eq:subsumption}
      $\myapprox$ $e_{1}$ \hfill {\sc subsumption}
\item $f(e_1,\ldots,e_m) \cap g(d_1,\ldots,d_n)$ \label{eq:failcons} 
      $\myapprox$ $\emptyset$
      ~if $f \not \equiv g$ or $n \not \equiv m$  \hfill {\sc clash} 
\item $f(e_1,\ldots,e_n) \cap f(d_1,\ldots,d_n)$ \label{eq:capcons}
      $\myapprox$ $f(y_1,\ldots,y_n)$      \hfill {\sc inter. distribution}  \\ ~if 
      $\forall j. 1 \leq j \leq n. y_j=e_j\cap d_j$
 
\item $f(y_1,\ldots,y_n)$ $\myapprox$ $\emptyset$ \label{eq:emptycons}
      ~if $\exists i.1\leq i\leq n. E\entails y_i = \emptyset$ \hfill {\sc emptiness}
\end{enumerate}

Note that line~\ref{eq:emptycons} makes use of the check
$E\entails y_i = \emptyset$. For this test a straightforward
adaptation of the type emptiness test of~\cite{Dart-Zobel}
to parameterized definitions is used.
Note also that the rewriting in line~\ref{eq:capcons} deserves some
explanation.  If the set equation system $E$ does not contain an
equation $y_j=e_j\cap d_j$ for some $j$ then the equation is added to
$E$, with $y_j$ a new fresh variable. Otherwise, $y_j$ from the
existing equation is used.
If equations were not added, it would prevent full normalization of
the set expressions. If variables were not reused, but instead 
new variables added each time, it would prevent the global algorithm from
terminating.

\paragraph{\textbf{Solving recurrence equations}.}
Since the set equation system is kept in standard form at all times,
the number of different equations that need to be considered is very
small. The procedure CASE reduces a given equation into a simpler
one. It basically takes care only of recurrences.
We call \emph{recurrence} an equation $x=e$ where $x$ occurs
top-level in $e$.  There are four cases of recurrences that are dealt
with: 
\begin{enumerate}
\item $x = x \myapprox x=\emptyset$
\item $x = x\cap e \leftrightarrow x \subseteq e \myapprox x=\emptyset$
\item $x = x\cup e \leftrightarrow e \subseteq x \myapprox x=e$
\item $x = (x\cap e_{1})\cup e_{2} \leftrightarrow e_{2} \subseteq x
  \subseteq e_{1} \myapprox x = e_{2}$
\end{enumerate}

Note that in all cases we have chosen for variable $x$ the least solution
of all possible ones allowed by the corresponding recurrence. We are thus
taking a minimal sufficient solution, in the sense that the resulting set of
terms would be the smallest possible one that still approximates the
program types.
This is possible because we keep equations in standard form, so that
their lhs is always a variable and there is only one equation per
variable. If the equation turns into a recurrence, then the program
contains a recursion which does not produce solutions (e.g., an
infinite failure).

\begin{example}\label{ex:recurrences}
Consider the following program with signatures \texttt{p(P)}, \texttt{q(Q)},
and \texttt{r(R)}:
\begin{verbatim}
p(X):- p(X).         q(a).                r(b).
                     q(Y):- q(Y).         r(Z):- q(Z), r(Z).
\end{verbatim}
It is easy to see that its initial set equation system
will progress towards the following form:
$ \{\ \texttt{P}=\texttt{P},\ \texttt{Q}=\texttt{a}\cup\texttt{Q},\ 
       \texttt{R}=\texttt{b}\cup (\texttt{Q}\cap\texttt{R})\ \}
$.
\end{example}

\begin{small}
\begin{algorithm}[t!]
\caption{SOLVE}
\label{algo:solve}
\algorithmicinput a set equation system $E$ in standard form. \\
\algorithmicoutput a set equation system $S$ in solved form.
\begin{algorithmic}[1]
\STATE initialize $S$
\REPEAT\label{solve:begin-whole}

  \STATE initialize $C$
  \REPEAT\label{solve:begin-main}
     \STATE subtract $q$ from $E$\label{solve:main-substract}
     \STATE $\langle q',C \rangle \leftarrow$  SIMP(DNF($q$),$C$,$S$)\label{solve:simp-1}
     \STATE $q'' \leftarrow$ CASE($q'$,$S$) 
     \STATE let $q''$ be of the form $x=e$
     \STATE replace every top-level occurrence of $x$ in $E$ and in $S$ by $e$\label{solve:replace-1} 
     \STATE add $q''$ to $S$
  \UNTIL{$E$ is empty}\label{solve:end-main}

  \REPEAT\label{solve:begin-emptiness}
  \FORALL{$q\in S$}
     \STATE $\langle q,C \rangle \leftarrow$ SIMP(DNF($q$),$C$,$S$)\label{solve:simp-2} 
     \IF{$q$ is of the form $x=e$, $e\neq\emptyset$, and $S\entails x=\emptyset$}\label{line:entails}
         \STATE replace $q$ by $x=\emptyset$ in $S$
     \ENDIF
  \ENDFOR
  \UNTIL{no equation $q$ is replaced}\label{solve:end-emptiness}

  \FORALL{$(x=\emptyset)\in S$}\label{solve:begin-propagate-fail}
     \IF{$x$ occurs in $Eq(P)$ for some clause $c$}
        \FORALL{variable $y$ in the head of $c$}
           \STATE replace the equation in $S$ with lhs $y$ by $y=\emptyset$\label{solve:propagate-fail}
        \ENDFOR
     \ENDIF
  \ENDFOR\label{solve:end-propagate-fail}

  \FORALL{$(x=e)\in S$}\label{solve:begin-update-C}
     \STATE replace every top-level occurrence of $x$ in $C$ by $e$ \label{solve:replace-2} 
  \ENDFOR\label{solve:end-update-C}
  \STATE add $C$ to $E$

  \IF{$E$ is empty} 
     \STATE $E \leftarrow$ BIND($S$)\label{solve:assign}
  \ENDIF

\UNTIL{$E$ is empty}\label{solve:end-whole}
\end{algorithmic}
\end{algorithm}
\end{small}

\vspace*{-3mm}
\paragraph{\textbf{The global algorithm}.}
The global algorithm for solving a set equation system is given as
Algorithm SOLVE.
In order to facilitate the presentation, the set equation
system is partitioned into three: $E$, $C$, and $S$. $E$ is the initial
equation system (generated from the program as in Eq.~\ref{eq:gen}),
$S$ is the solved system (i.e., the output of the analysis), and $C$ is an
auxiliary system with the same form as $E$.
This forces the use of SIMP in lines~\ref{solve:simp-1}
and~\ref{solve:simp-2} with two input equation systems instead of one.
SIMP may add new equations, and these are added to $C$. Also, in SIMP
the test in line~\ref{eq:emptycons} ($E\entails y_j=\emptyset$) is
carried along in $S$ (i.e., it should be read as $S\entails
y_j=\emptyset$ when called from SOLVE).

Once more, we use the test $S\entails x=\emptyset$
(line~\ref{line:entails} of Algorithm SOLVE). This is related to the
other use of this test in SIMP. Note that SIMP is invoked again at
line~\ref{solve:simp-2} of Algorithm SOLVE. Together, the loop in
lines~\ref{solve:begin-emptiness}-\ref{solve:end-emptiness} of
Algorithm SOLVE and line~\ref{eq:emptycons} of SIMP perform a complete
type emptiness propagation.  Because of this, the effect of the rest
of cases of SIMP, which propagate symbol $\emptyset$ in
subexpressions, is to achieve an implicit check for non-emptiness of
intersections.
The rationale behind the loop in
lines~\ref{solve:begin-propagate-fail}-\ref{solve:end-propagate-fail} is 
different. It solves a lack of propagation of failure for some programs,
due to the form of the initial equation system.

\begin{example}\label{ex:propagate-empty}
In the program below, failure of \texttt{p/1} would not be detected unless
step~\ref{solve:propagate-fail} is performed, even if failure of \texttt{q/2}
is detected because of type emptiness.
\begin{verbatim}
p(X):- q(b,X).         q(a,a).
\end{verbatim}
\end{example}

\comment{For p(a):- q(b). q(a). we need to guarantee that there is always
  a head variable on which to hang for the propagation. Another program
  transformation is needed: every head arg should have at least one var.
}

Finally, the BIND procedure invoked at line~\ref{solve:assign} of
Algorithm SOLVE is not needed for obtaining the solved equation
system, since $S$ is already in solved form when BIND is called, but
it may improve the precision of the analysis.

\begin{example}[SOLVE with binding of free variables]
\label{ex:analysis}
We show the analysis of a contrived example which exposes the
strength of our approach in propagating types through the type
variables that act as parameters.
We consider signature \texttt{append(A1,A2,A3)} already solved with
solution:
\texttt{A1 = []} $\cup$ \texttt{[X|A1]},
\texttt{A3 = A2} $\cup$ \texttt{[X|A3]};
and concentrate on the analysis of: 
\begin{verbatim}
appself(A,B):- append(A,[],B).
\end{verbatim}

\noindent
We show the contents of $E$ at the beginning of Algorithm SOLVE and of
$S$ at line~\ref{solve:assign} ($C$ and $E$ are empty).  We only show
the equations for the program variables of \texttt{appself/2}. The
equations missing are those for the solution of \texttt{append/3}
--note that \texttt{A2} is free-- and aliases of variables \texttt{A}
and \texttt{B} for the arguments of the signature of
\texttt{appself/2}:
\\[-0.5\baselineskip]

\noindent
\begin{tabular}{l}
E=\{ \texttt{A=A1}, \texttt{B=A3}, \texttt{W=A2} $\cap$ \texttt{[]} \}
\\
S=\{ \texttt{A=[]} $\cup$ \texttt{[X|A1]}, \texttt{B=A2} $\cup$ \texttt{[X|A3]}, 
     \texttt{W=A2} $\cap$ \texttt{[]} \}
\end{tabular}
\\

\noindent
Now, BIND can give a more precise value to the
only free variable that appears in an intersection, \texttt{A2} 
(\texttt{A2} = \texttt{[]}).
We now include the equation for \texttt{A3}, which is relevant:
\\[-0.5\baselineskip]

\noindent
\begin{tabular}{l}
E=\{\texttt{A2=[]}\}
\\
S=\{\texttt{A=[]} $\cup$ \texttt{[X|A1]},~ \texttt{B=A2} $\cup$ \texttt{[X|A3]},~
    \texttt{W=A2} $\cap$ \texttt{[]},~ \texttt{A3=A2} $\cup$ \texttt{[X|A3]}\} \\
\end{tabular}
\\[-0.5\baselineskip]

\noindent
After a second iteration of the main
loop (lines~\ref{solve:begin-whole}--\ref{solve:end-whole}) we have
$E$ empty again, and:
\\[-0.5\baselineskip]

\noindent
\begin{tabular}{l}
S=\{ \texttt{A=[]} $\cup$ \texttt{[X|A1]},~ 
     \texttt{B=[]} $\cup$ \texttt{[X|A3]},~
     \texttt{W=[]},~ \texttt{A3=[]} $\cup$ \texttt{[X|A3]} \} \\
\end{tabular}
\\[-0.5\baselineskip]

\noindent
Finally, after projection on the variables of interest (which are in
fact those of the signature of \texttt{appself/2},
but we use its program variables \texttt{A},\texttt{B} for clarity
instead): 
\\[-0.5\baselineskip]

\noindent
\begin{tabular}{l}
S=\{ \texttt{A=[]} $\cup$ \texttt{[X|A]},~
     \texttt{B=[]} $\cup$ \texttt{[X|B]} \} \\
\end{tabular}
\end{example}

\paragraph{\textbf{Binding free variables to values}.}
Since the equation system is already solved, i.e., it is in solved form,
when BIND is applied, free variables can at this moment take any value.
BIND will then mimic unification by binding free variables to the
minimal values required so that the expressions involving them have a
solution. In order to do this, all subexpressions with the form of an
intersection (i.e., unification) with a free variable are considered.
These are called \emph{conjuncts}.

Since the set expressions are in parameterized form,
conjuncts are all of the form $e_1\cap e_2$, where $e_1$ is a,
possibly singleton, conjunction of free variables, and $e_2$ a,
possibly non-existing, constructor expression. Let $e_1$ be of the
form $x_1\cap\ldots\cap x_n$, $n\geq 1$, and $e$ be $e_2$ if it
exists or a new variable, otherwise. A set of candidates is proposed
of the form $\{ x_i=e ~|~ 1\leq i\leq n \}$. For an equation $q$,
let $Candidates(q)$ denote the set of sets of candidates for each
conjunct in $q$.
For defining procedure BIND we need to consider the following relation
between equations. Let $q\mapsto q'$ when equation $q$ is the equation
taken by SIMP at a given time, and $q'$ the new equation added at
step~\ref{eq:capcons} of Algorithm SIMP from $q$.  Let $\mapsto^*$
denote the transitive closure of $\mapsto$.  BIND constructs the
formula below, and synthesizes from it a suitable set of set
equations.
%
\begin{equation}
\label{eq:bind}
\bigwedge_{q\in Eq(P)} ~~ \bigvee_{q\mapsto^* q'} ~~ 
\bigvee_{c\in Candidates(q')} ~~ \bigwedge_{e\in c} e
\end{equation}

\begin{example}
Let the following contrived program for \texttt{alternate/2}, which
resembles the way the classical program for the towers of Hanoi problem
alternates the arguments representing the pegs that hold the disks
across recursions:
\begin{verbatim}
alternate(A1,B1).
alternate(A2,B2):- alternate(B2,A2).
\end{verbatim}
It is easy to see that, for a signature \texttt{alternate(P1,P2)}, 
the solution will be the system of the following two equations:
\\[-0.5\baselineskip]

\noindent
\begin{tabular}{ll}
& \texttt{P1 = A1} $\cup$ \texttt{B1},
~~~~
  \texttt{P2 = A1} $\cup$ \texttt{B1}
\end{tabular}
\\[-0.5\baselineskip]

The analysis of a program containing an atom of the form 
\texttt{alternate(a,b)} in a clause body, will be faced with
the equations:
\\[-0.5\baselineskip]

\noindent
\begin{tabular}{ll}
& $\texttt{W1 = } ( \texttt{A1} \cup \texttt{B1} ) \cap \texttt{a}$,
~~~~
  $\texttt{W2 = } ( \texttt{A1} \cup \texttt{B1} ) \cap \texttt{b}$,
\end{tabular}
\\[-0.5\baselineskip]

\noindent
which at line~\ref{solve:assign} of Algorithm SOLVE would be solved
as:
\\[-0.5\baselineskip]

\noindent
\begin{tabular}{ll}
& $\texttt{W1 = } ( \texttt{A1} \cap \texttt{a} )\cup 
                  ( \texttt{B1} \cap \texttt{a} )$,
~~~~
  $\texttt{W2 = } ( \texttt{A1} \cap \texttt{b} )\cup 
                  ( \texttt{B1} \cap \texttt{b} )$,
\end{tabular}
\\[-0.5\baselineskip]

\noindent
so that BIND will find two original equations (in $Eq(P)$; not related
by $\mapsto^*$) with two candidates each 
($\{\{\texttt{A1=a}\},\{\texttt{B1=a}\}\}$
for the equation for \texttt{W1} and
$\{\{\texttt{A1=b}\},\{\texttt{B1=b}\}\}$
for the equation for \texttt{W2}).
Thus, BIND obtains the formula:
\\[-0.5\baselineskip]

\noindent
\begin{tabular}{ll}
& $(\ \texttt{A1 = a}\ \vee\ \texttt{B1 = a}\ )\ \wedge\
   (\ \texttt{A1 = b}\ \vee\ \texttt{B1 = b}\ )$,
~~~~which is equivalent to:
\end{tabular}
\\[-0.5\baselineskip]

\noindent
\begin{tabular}{ll}
& $(\ \texttt{A1 = a}\ \wedge\ \texttt{B1 = b}\ )\ \vee\
   (\ \texttt{A1 = b}\ \wedge\ \texttt{B1 = a}\ )$,
~~~~and results in the following two
\end{tabular}
\\[-0.5\baselineskip]

\noindent
set equations, which ``cover'' all solutions of the above formula:
\\[-0.5\baselineskip]

\noindent
\begin{tabular}{ll}
& \texttt{A1 = a} $\cup$ \texttt{b},
~~~~\texttt{B1 = a} $\cup$ \texttt{b},
\end{tabular}
\\[-0.5\baselineskip]

\noindent
and (imply expressions for \texttt{P1} and \texttt{P2} which)
correctly approximate the success types of \texttt{alternate/2}
when called as in \texttt{alternate(a,b)}.
\end{example}

In the synthesis of set equations from the formula in Eq.~\ref{eq:bind},
$\vee$ turns into $\cup$ and $\wedge$ turns into $\cap$. The details of
how to do this in a complete synthesis procedure are the subject of
current work. Our
implementation currently deals with the simpler cases, including those
similar to the example above and those in which all candidates involve
only one free variable (as in Ex.~\ref{ex:analysis}).

\begin{small}
\begin{table}[t]
\begin{center}
\begin{tabular}{|c|c|c|c|c|c|c|c|c|} 
\hline
             &    &   \multicolumn{3}{c|}{\textsf{NFTA}} & 
                      \multicolumn{3}{c|}{\textsf{PTA}} &  \\ 
\cline{3-8}
\textsf{Benchmark} & 
\textsf{\# Desc.} &
\textsf{Prec.} & 
\textsf{Time (s)} & 
\textsf{Error} & 
\textsf{Prec.} & 
\textsf{Time (s)} & 
\textsf{Error} & 
\textsf{Bad call type} \\
\hline
\hline
 \textsf{append} & 3 & 1 & 0.000 & n & 3 &  0.000 & \textsf{y} & \texttt{append(A,a,A)} \\ 
 \hline  
 \textsf{blanchet} & 1 & 1 & 0.052 & \textsf{y} & 1  &  0.032 & \textsf{y} & \texttt{attacker(s)} \\ 
 \hline  
 \textsf{dnf} & 2  & 2  & 1.156 & n & 2 &  1.892 & n & \texttt{dnf(X,a(z1, o(z2,z3)))} \\ 
 \hline  
 \textsf{fib} & 2 & 0 & 0.000 & n & 2  &  0.000 & \textsf{y} & \texttt{fib(a,X)} \\ 
 \hline  
 \textsf{grammar} & 1 & 0 & 0.016 & \textsf{y} & 1  &  0.124 & \textsf{y} & \texttt{parse([boxes,fly],S)}\\ 
 \hline  
 \textsf{hanoi} & 5 & 1  & 0.020 & n & 5  &  0.028 & n & \texttt{hanoi(5,a,b,c,[mv(e,f)])}\\ 
 \hline  
 \textsf{mmatrix} & 3 & 3 & 0.008 & n & 3  &  0.008 & \textsf{y} & \texttt{mmult([1,2],[[1,2],[3,4]],X)}\\ 
 \hline  
 \textsf{mv} & 3  & 1  & 0.020 & n & 3  &  0.036 & \textsf{y} & \texttt{mv([1,3,1],[b,c,a],X)}\\ 
 \hline  
 \textsf{pvgabriel} & 2 & 0 & 0.184 & n  & 2  &  0.124 & n  & \texttt{pv\_init([1,2],X)} \\ 
 \hline  
 \textsf{pvqueen} & 2 & 1 & 0.020 & n & 2 &  0.028 & \textsf{y} & \texttt{queens(4,[a,b,c,d])}\\ 
 \hline  
 \textsf{revapp} & 2  & 2 & 0.004 & n & 2 &  0.004 & \textsf{y} & \texttt{rev([1,2],[a,b])} \\ 
 \hline  
 \textsf{serialize} & 2 & 2 & 0.044 & n & 2 &  0.164 & \textsf{y} & \texttt{serialize(``hello'',[a,b,c])}\\ 
 \hline  
 \textsf{zebra} & 7  & 1 & 0.080 & n & 7  &  0.372 & \textsf{y} & \texttt{zebra(E,S,J,U,second,Z,W)}\\ 
 \hline  
 \hline
 \textsf{Total} & 35 & 13 & 1.604   & 2 & 35 & 2.812  & 10 & \\
 \hline  
\end{tabular}
\end{center}
\vspace{-0.5\baselineskip}
\caption{Experimental results for NFTA and parameterized type analysis (PTA)}
\label{table:results}
\vspace{-\baselineskip}
\end{table}
\end{small}

\section{Preliminary experimental evaluation}
\label{sec:results}

In order to study the practicality of our method we have implemented
a prototype 
analyzer in \ciao\ 
(\texttt{http://www.ciaohome.org}~\cite{ciao-reference-manual-1.13-short})
and processed a representative set of benchmarks taken mostly from
the PLAI (the \ciao\ program analyzer) and
GAIA~\cite{VanHentenryckCortesiLeCharlier94} sets.
We chose the \emph{Non-deterministic Finite Tree Automaton}
(NFTA)-based analysis~\cite{set-based-absint-padl-short} for 
comparison. 
We believe that this 
is a fair comparison since
it also over-approximates the success set of a program in a bottom-up
fashion (inferring regular types). 
Its implementation is publicly available 
(\texttt{http://saft.ruc.dk/Tattoo/index.php}) and it is
also written in \ciao.
We decided not to compare herein with top-down, widening-based
type analyses 
(e.g.~\cite{janss92,VanHentenryckCortesiLeCharlier94,eterms-sas02}),
since it is not clear how they relate 
to our method. We 
leave this comparison as interesting future work.

The results are shown in Table~\ref{table:results}. The fourth and
seventh columns show
analysis times in seconds for both analyses on
an Intel Core Duo 1.33GHz CPU, 1GB RAM, and Ubuntu 8.10 Linux OS. The time
for reading the program and generating the constraints is omitted
because it is always negligible compared to the analysis time. 
Column \textsf{\# Desc.} shows the number of type descriptors
(i.e., predicate argument positions) which will be considered for the
accuracy test (all type descriptors are considered for the timing
results).  For simplicity, we report only on the argument positions
belonging to the main predicate. Note that execution of these
predicates often reaches
all the predicates in the program and thus the accuracy of the types
for those positions is often a good summary of precision for the whole
program. 
Columns labeled \textsf{Prec.} show the number of
type descriptions inferred by NFTA and our approach that are different
from type \emph{any}.
To test precision further, we have added to each benchmark a query that
fails, which is shown in the last column (\textsf{Bad call type}).
Columns labeled \textsf{Error} show whether this is captured by
the analysis or not.

Regarding \emph{accuracy}, 
our experiments show that parameterized types allow inferring type
descriptors with significant better precision. Our approach inferred
type descriptions different from \emph{any} for every one of the 35
argument positions considered, while NFTA inferred only 13. Moreover,
those type descriptors were accurate enough to capture type emptiness
(i.e., failure) in 10 over 13 cases,
whereas NFTA could only catch two errors.
The cases of \textsf{dnf}, \textsf{hanoi}, and
\textsf{pvgabriel} deserve special attention, since our method could not capture
emptiness for them. For \textsf{dnf} the types inferred are not precise enough
to capture that the second argument is not in disjunctive normal
form. %
This is due to the lack of 
inter-variable dependencies in our set-based approach. However,
emptiness  is easily captured for other
calls such as \texttt{dnf(X,b).} A similar
reasoning holds for \textsf{hanoi} since the analysis infers that the types
of the pegs are the union of \texttt{a}, \texttt{b}, \texttt{c},
\texttt{e}, and \texttt{f}. Finally, the success of \textsf{pvgabriel}
depends on a run-time condition, and thus no static analysis can catch
the possible error.

Regarding \emph{efficiency}, we expected our analysis to be slower
than other less expressive methods (like NFTA), since our method is more
expressive than previous proposals. Table~\ref{table:results} shows that 
this is indeed the case, but the 
differences are reasonable for the selected set of benchmarks,
specially considering the improvement in accuracy.
Further research %
is of course needed using larger programs (see
Sect.~\ref{sec:conclusionsfuture}), but we find the results clearly
encouraging.  

\section{Related work}
\label{sec:related}

Type inference, i.e., inferring type definitions from a
program, has received a lot of attention in logic programming.
Mishra and Reddy~\cite{Mishra84-short,Mishra85} propose the inference of
ground regular trees that represent types, and compute an upper
approximation of the success set of a program. The types inferred are
monomorphic.
Zobel~\cite{zobel87-short} also proposes a type inference method for a program, where
the type of a logic program is defined as a recursive (regular) superset
of its logical consequences. However, the inference procedure does not derive
truly polymorphic types: the type variables are just names for types
that are defined by particular type rules.

In~\cite{LakshmanReddy91} an inference method called type reconstruction
is defined,
which derives types for the
predicates and variables of a program. The types are polymorphic
but they are fixed in advance. On the contrary, our inference method 
constructs type definitions during the analysis.
In~\cite{FruewirthShapiroVardiYardeni91} the idea is
that the least set-based model of a logic program can be 
seen as the exact Herbrand model of an approximate logic program. This
approximate program is a regular unary program, which has a specific
syntactic form which limits its computational power. For example, type
parameters cannot be expressed.
The work of~\cite{gallagher-types-iclp94,Saglam-Gallagher-94}
also builds regular unary programs, and their type inference derives
such programs. However, the types remain monomorphic.

Heintze and Jaffar~\cite{HeintzeJaffar90} defined an elegant method for
semantic approximation, which was the origin of set-based analysis of
logic programs. The semantics computes, using set substitutions, 
a finite representation of a model of the program that is
an approximation of its least model. Unfortunately,
the algorithm of~\cite{HeintzeJaffar90} was rather complicated and
practical aspects were not 
addressed. Such practical aspects 
were addressed instead in subsequent
work~\cite{Heintze92,HeintzeJ92b,Heintze92PhD}, simplifying the
process. However, our equations 
are still simpler: for example, we do not make use of \emph{projections},
i.e., expressions of the form $f^i(x)$, where $f$ is a constructor,
the meaning of, e.g., $y=f^1(x)$ being that $x\supseteq f(y,\ldots)$.
Also, and more importantly, these approaches do not take advantage of
parameters, and the types obtained are again always monomorphic.

Type inference has also been approached using the technique of abstract
interpretation. A fixpoint is computed, in most cases with the aid 
of a widening operation to limit the growth of the type domain. 
This technique is also used in~\cite{gallagher-types-iclp94}.
However, none of the analyses proposed to date 
(e.g.,~\cite{janss92,VanHentenryckCortesiLeCharlier94,eterms-sas02})
is parametric.
It is not clear how the approach based on abstract domains with widening
compares to the set constraints-based approach. An interesting avenue
for future work, however, is to define our analysis as a fixpoint. 
This is possible in general due to a result by Cousot and
Cousot~\cite{CousotC95-short} 
which shows that rewriting systems like ours can be defined alternatively
in terms of a fixpoint computation.

Directional types~\cite{Aiken94directionaltype,Cha00} are based on
viewing a predicate as a procedure that maps a call type
to a success type. This captures some dependencies between arguments,
but they are restricted to monomorphic types.
It would be interesting to build this idea into our approach.
We believe that some imprecision found in analyzing recursive calls
with input arguments 
could be alleviated by using equations which captured both calls and 
successes. The relation of these with directional types might be
worth investigating.

Bruynooghe, Gallagher, Humbeeck, and Schrijvers~\cite{BGV05,SBG08} 
develop analyses for type inference which are also based on the set
constraints approach. However, their analyses infer well-typings, so
that the result is not an approximation of the program success set,
as in our case. In consequence, their algorithm is simpler, mainly
because they do not need to deal with intersection. In~\cite{SBG08}
the monomorphic analysis of~\cite{BGV05} is extended to the
polymorphic case, using types with an expressiveness comparable to
our parameterized types. 

The set constraints approach to type analysis is also taken 
in NFTA~\cite{set-based-absint-padl-short}, which is probably the closest 
to ours. However, the types inferred in this analysis,
which are approximations of the program success set, are not
parametric and as a result the analysis is less precise than ours,
as shown in our experiments.
The work of~\cite{bruy-gallag-lopstr04} aims at inferring parametric
types that approximate the success set. Although independently
developed, it includes some of the ideas we propose. However,
the authors do not use the set-based approach, and their inference algorithm is
rather complex. The latter made the authors 
abandon that line of work and the approach was not developed, 
switching instead to well-typings.

\section{Concluding remarks and future work}
\label{sec:conclusionsfuture}

To conclude, using the set constraints approach we have proposed a
simple type inference based on the rewriting of equations which,
despite its simplicity, allows enhanced expressiveness. This is
achieved by using type variables with a global scope as true
parameters of the equations. The improved expressiveness allows for
better precision, at an additional cost in efficiency, which is,
nevertheless, not high.

During our tests, we have identified
potential bottlenecks in our analysis which may appear in large
programs and which are worth investigating in future research.
Some practical issues were already addressed by Heintze in~\cite{Heintze92},
that we have not considered, concentrating instead in this paper 
on the precision and soundness of our approach.
%
In particular, it is well-known that a naive DNF expansion
may make the size of an expression grow exponentially. However, we use
simple rules for minimizing the number of expressions (such as
computing only intersections that will survive after simplification)
which work quite well in practice. Even for larger programs we expect
these rules to be effective, based on the experience of~\cite{Heintze92}.
%
An efficient method for storing the new intersections generated in
line~\ref{eq:capcons} in SIMP is also needed for the scalability of the system. We
use the same simple technique as~\cite{Heintze92}. Note that the size
of the table is in the worst case $2^{N}$ where $N$ is the number of
original variables in the program.
However, the size of the table actually
grows almost linearly in our experiments (we omitted these results due to
space limitations). Even if the size of the table were to grow faster,
we believe that we can mitigate this effect with similar techniques
(e.g., use of BDDs) to~\cite{set-based-absint-padl-short,GalHenrBanda05},
because of the high level of redundancy.  
%
We have also observed that the replacement
performed in lines~\ref{solve:replace-1} and~\ref{solve:replace-2} 
of SOLVE may be expensive for large programs.
We think that there are many opportunities 
for reducing this limitation (e.g., dependency directed
updating~\cite{Heintze92}).

{\small

\ \\
\textbf{Acknowledgments:} The authors would like to thank John
Gallagher and the anonymous referees for useful comments on previous
versions of this paper.  This work was funded in part by EU projects
FET IST-15905 {\em MOBIUS}, IST-215483 {\em SCUBE}, FET IST-231620
{\em HATS}, and 06042-ESPASS, Ministry of Science projects
1TIN-2008-05624 {\em DOVES}, Ministry of Industry project
FIT-340005-2007-14, CAM project S-0505/TIC/0407 {\em PROMESAS},
and Singapore Ministry of Education Academic Research Fund
No. RP-252-000-234-112.

\bibliographystyle{plain}

}

\end{document}